\documentclass[pre,showpacs,showkeys,floatfix,nofootinbib,longbibliography,onecolumn]{revtex4-1}

\usepackage{graphicx}
\usepackage{amsmath}
\usepackage{amsfonts}
\usepackage{bm}
\usepackage{bbm}
\usepackage{amssymb}
\usepackage{mathrsfs}
\usepackage{dsfont}
\usepackage{subfigure}

\newlength{\irrl}
\newlength{\irrw}
\newcommand{\irr}[1]{
  \settowidth{\irrl}{\mbox{$\displaystyle #1$}}
  \setlength{\irrw}{0.12ex}
    \mbox{$\hspace{0.2em}
    \stackrel{
      \mbox{$\vphantom{\rule[-5\irrw]{\irrw}{6\irrw}}
             \rule[-4\irrw]{\irrw}{5\irrw}\hspace{-\irrw}
             \rule{\irrl}{\irrw}\hspace{-\irrw}
             \rule[-4\irrw]{\irrw}{5\irrw}$}}
     {\mbox{$\displaystyle #1$}}\hspace{0.2em}$}
}

\newcommand{\p}{\bm{p}}
\newcommand{\PP}{\mathbf{P}}

\newcommand{\avir}[1]{\left\langle\irr{{#1}}\right\rangle}
\newcommand{\avirt}[2]{\left\langle\irr{{#1}}\right\rangle_{#2}}
\newcommand{\avet}[2]{\left\langle{#1}\right\rangle_{#2}}
\newcommand{\aved}[1]{\avet{#1}{\dens}}
\newcommand{\avird}[1]{\avirt{#1}{\dens}}
\newcommand{\ave}[1]{\left\langle{#1}\right\rangle}
\newcommand{\Pn}[1]{\PP^{({#1})}}
\newcommand{\pn}[1]{\p^{\otimes{#1}}}

\newcommand{\length}[1]{s({#1})}

\newcommand{\A}{\mathbf{A}}
\newcommand{\B}{\mathbf{B}}
\newcommand{\C}{\mathbf{C}}

\newcommand{\va}{\bm{a}}

\newcommand{\vd}{\bm{d}}
\newcommand{\e}{\bm{e}}
\newcommand{\x}{\bm{x}}
\newcommand{\etwo}{\e\otimes\e}

\newcommand{\ptwo}{\p\otimes\p}

\newcommand{\athree}{\va\otimes\va\otimes\va}

\newcommand{\pthree}{\p\otimes\p\otimes\p}

\newcommand{\pj}{\p^{\otimes j}}

\renewcommand{\frame}{(\e_1,\e_2)}

\newcommand{\disk}{{\mathbb{S}^1}}

\newcommand{\real}{{\mathbb{R}}}

\newcommand{\dens}{\varrho}

\newcommand{\frax}[2]{\textstyle\frac{#1}{#2}}

\begin{document}
\title{Octupolar order in two dimensions}
\author{Epifanio G. \surname{Virga}}
\email[e-mail: ]{eg.virga@unipv.it}
\affiliation{Dipartimento di Matematica, Universit\`a di Pavia, Via Ferrata 5, I-27100 Pavia, Italy}

\date{\today}

\begin{abstract}
Octupolar order is described in two space dimensions in terms of the maxima (and conjugated minima) of the probability density associated with a third-rank, fully symmetric and traceless tensor. Such a representation is shown to be equivalent to diagonalizing the relevant third-rank tensor, an equivalence which however is only valid in the two-dimensional case.
\end{abstract}



\maketitle

\section{Introduction}\label{sec:intro}
Octupolar and tetrahedratic are synonymous adjectives when applied to soft matter ordering.\footnote{Despite their equivalence, we shall use throughout the former instead of the latter.} Loosely speaking, they are called upon whenever a third-rank tensor is needed to describe order in a molecular ensemble. Perhaps, Fel~\cite{fel:tetrahedral,fel:symmetry} was the first to consider a third-rank tensor to describe an unconventional liquid-crystal phase with the tetrahedron symmetry condensing directly from the isotropic phase.\footnote{A wealth of unconventional nematic phases allowed by symmetry are described in \cite{mettout:macroscopic}. Fel's analysis was revisited in \cite{radzihovsky:fluctuation}.} While these early studies had concerned phases fully characterized by a third-rank tensor, a subsequent, comprehensive analysis considered the coexistence of polar, quadrupolar, and octupolar orders, elucidating the complex network of phase transitions  thus made possible \cite{lubensky:theory}.
Non-polar nematic phases, where only quadrupolar and octupolar orders can coexist, were studied in \cite{brand:macroscopic}, where the reader is also referred to for a rather accurate and informative review of earlier contributions to theory.\footnote{A full macroscopic theory is also proposed in \cite{brand:macroscopic}, which encompasses statics and hydrodynamics of these phases, which are referred to as $D2d$ phases. In particular, a macroscopic dynamic theory for the octupolar $T_d$ phase is presented in Sect.~2.2 of \cite{brand:macroscopic}, building on an earlier work \cite{brand:flow}.} Third-rank tensors do not only play a role in describing new condensed soft phases; they have also recently been employed in the active dynamics of self-propelled microorganisms. In a series of papers \cite{ohta:deformation,hiraiwa:dynamics,tarama:oscillatory}, extending the ideas originally presented in \cite{ohta:deformable}, a third-rank tensor is invoked to represent the asymmetric shape of a living cell when spontaneous deformation and drift velocity are intimately interconnected.

Here our attention is mainly directed to describing octupolar order through a third-rank, fully symmetric and traceless tensor. It is known that in three space dimensions that are $7$ independent such tensors \cite{lubensky:theory}; they become $2$ in two space dimensions. To diagonalize a third-rank tensor is not a univocally defined task, as no analogue of the Spectral Theorem exists for symmetric tensors with rank higher than $2$. Likewise, a commonly accepted notion of generalized eigenvalues and eigenvectors has not been established yet for these tensors, which would univocally identify the former as scalar order parameters.\footnote{This issue has already been addressed in a previous study \cite{zheng:eigenvalue} which proposed an approach alternative to the one followed here. Likewise, other computational definitions of scalar order parameters for both tetrahedral and cubatic symmetries can also be found in \cite{romano:computer:06,romano:computer}.} We shall find a viable substitute for such a classical representation of the scalar order parameters in the maxima (and conjugated minima) of the octupolar probability density that shall be introduced in Sec.~\ref{sec:density_multipoles}.

In this paper, only the two-dimensional case will be considered: it is far easier than the three-dimensional case (which will be studied elsewhere \cite{gaeta:octupolar}), but not unrealistic, as suggested by recent experiments with nails on a vibrating table \cite{macdonald:pattern,morris:movie}. Compared to either tetrahedratic phases or self-propelling microorganisms, this is perhaps a more mundane manifestation of the need for an octupolar order descriptor. Section~\ref{sec:dipole_and_quadrupole} is a brief interlude with the purpose of reinterpreting the traditional descriptions of both dipolar and quadrupolar orders in terms of maxima (and conjugated minima) of the appropriate probability density. In Sec.~\ref{sec:octupole}, which is the heart of the paper, the octupolar order in two space dimensions is formally characterized in terms of the generalized eigenvalues and eigenvectors of a third-rank tensor, which is also given an equivalent diagonal representation, whose validity is however restricted to the two-dimensional case. In Sec.~\ref{sec:conclusion}, both strategy and conclusion of this paper are recapitulated and their foreseeable extensions to the three-dimensional case are briefly anticipated.

\section{Probability density multipoles}\label{sec:density_multipoles}
\begin{figure}[h]
\centering
\includegraphics[width=0.15\linewidth]{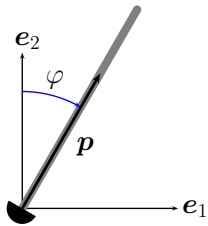}
\caption{(Color online) Cartoon illustrating a nail and the polar unit vector $\p$ represented as in \eqref{eq:p_representation_phi} relative the Cartesian frame $\frame$.}
\label{fig:nail}
\end{figure}

Consider, for definiteness, a system of rigid \emph{nails} in a plane like those studied experimentally on a vibrating table \cite{macdonald:pattern,morris:movie}. Each nail is characterized by a unit vector $\p$ oriented from its head to its end (see Fig.~\ref{fig:nail}).
Assume that nails are distributed in the plane with a certain probability law which may induce order in their ensemble. The probability distribution density, which is defined on the unit circle $\disk$, will be denoted by $\dens:\disk\to\real^+$; it is subject to the normalization condition
\begin{equation}\label{eq:normalization_condition}
\int_\disk\dens(\p)d\length{\p}=1,
\end{equation}
where $\length{\p}$ is the arch-length on $\disk$. Adapting the formalism of \cite{turzi:cartesian} to represent Buckingham's formula \cite{buckingham:angular} for $\dens$ in terms of Cartesian tensors, we write
\begin{equation}\label{eq:probability_density_representation}
\dens(\p)=\frac{1}{2\pi}\left(1+\sum_{j=1}^\infty\Pn{j}\cdot\pj\right),
\end{equation}
where $\pj$ is the $j$-th rank tensor defined by
\begin{equation}\label{eq:p_j_definition}
\pj:=\underbrace{\p\otimes\dots\otimes\p}_{j\ \text{times}},
\end{equation}
$\otimes$ denotes tensor product, and $\Pn{j}$ is a $j$th-rank tensor soon to be related to the $j$th moment of $\dens$.\footnote{In general, $\dens$ could alternatively be represented as an expansion in symmetry-adapted Wigner rotation matrices \cite{zannoni:distribution}, but both here and in a forthcoming paper \cite{gaeta:octupolar} we are interested in the equivalent Cartesian tensor representation of $\dens$.} In \eqref{eq:probability_density_representation}, the inner product $\mathbf{A}^{(j)}\cdot\mathbf{B}^{(j)}$ of two tensors, $\mathbf{A}^{(j)}$ and $\mathbf{B}^{(j)}$ of equal rank $j$, corresponds to the following contraction of components in any Cartesian frame $\frame$ of the plane:
\begin{equation}\label{eq:inner_product_of_tensors}
\mathbf{A}^{(j)}\cdot\mathbf{B}^{(j)}:=A_{i_1\dots i_j}B_{i_1\dots i_j}.
\end{equation}
Here and in what follows we understand that repeated indices are summed over their whole range.

Since the products in \eqref{eq:p_j_definition} are completely symmetric under the exchange of any pair of components for all $j$, by \eqref{eq:inner_product_of_tensors} so are also required to be all $\Pn{j}$ to dispose of redundant components. For the function $\dens$ in \eqref{eq:probability_density_representation} to obey \eqref{eq:normalization_condition}, all $\Pn{j}$ are further required to be traceless in any pair of components, as isotropy of the plane demands that the tensor
\begin{equation}\label{eq:isotropic_average}
\avet{\pj}{0}:=\frac{1}{2\pi}\int_\disk\pj d\length{\p}
\end{equation}
either vanishes, if $j$ is odd, or its Cartesian components can be written as symmetrized products of Kronecker's $\delta$'s, if $j$ is even. Letting the brackets $\irr{\dots}$ denote the irreducible, completely symmetric and traceless part of any tensor they surmount, the above properties of each $\Pn{j}$ are embodied by the equation
\begin{equation}\label{eq:symmetries_of P_j}
\irr{\Pn{j}}=\Pn{j}\quad\forall\ j.
\end{equation}
We shall denote by $\avet{\dots}{\dens}$ the ensemble average relative to $\dens$:
\begin{equation}\label{eq:ensemble_average_definition}
\avet{\dots}{\dens}:=\int_\disk(\dots)\dens(\p)d\length{\p},
\end{equation}
so that the average $\avet{\dots}{0}$ in \eqref{eq:isotropic_average} corresponds to the average relative to the isotropic density function $\dens_0\equiv\frac{1}{2\pi}$.

It is a direct consequence of \eqref{eq:symmetries_of P_j} and \eqref{eq:probability_density_representation} that
\begin{equation}\label{eq:average_of_p_j}
\avet{\irr{\pj}}{\dens}=\avet{\pn{2j}}{0}\circ\Pn{j}\quad\forall\ j\geqq1,
\end{equation}
where $\circ$ denotes tensor multiplication.\footnote{If $A_{i_1\dots i_{j}i_{j+1}\dots i_{2j}}$ and $B_{h_1\dots h_j}$ are the Cartesian components of tensors $\A$ and $\B$, of rank $2j$ and $j$ respectively, then $\C=\A\circ\B$ is a tensor of rank $j$ and its components $C_{i_1\dots i_j}$ are given by $C_{i_i\dots i_j}=A_{i_1\dots i_jh_1\dots h_j}B_{h_1\dots h_j}$.} While it is almost immediate to prove from \eqref{eq:average_of_p_j} that
\begin{equation}\label{eq:P_1}
\aved{\p}=\frac12\Pn{1},
\end{equation}
it requires some tedious labour to show that
\begin{equation}\label{eq:P_2}
\avir{\pj}=\frac{1}{2^j}\Pn{j}\quad\forall\ j\geqq2,
\end{equation}
so that we can rewrite \eqref{eq:probability_density_representation} as
\begin{equation}\label{eq:probability_density_moments}
\dens(\p)=\frac{1}{2\pi}\left(1+2^j\avird{\pj}\cdot\pj\right).
\end{equation}
In \eqref{eq:probability_density_moments}, $\dens$ is expressed as the sum of \emph{density multipoles}, each associated with a corresponding order tensor.\footnote{Which represents density moments of the appropriate rank.} We are especially interested in the first three \emph{order tensors}, $\aved{\p}$, $\avird{\ptwo}$, and $\avird{\pthree}$, featuring in \eqref{eq:probability_density_moments}, which represent three independent descriptors of order; we call them \emph{dipole}, \emph{quadrupole}, and \emph{octupole}, respectively. The first and the last are measures of \emph{polarity}, the latter becoming relevant when the former is bound not to dominate, as suggested by the tendency of shacked nails shown in \cite{macdonald:pattern,morris:movie} to be on average combined in antiparallel pairs. For a pair of nails, it was indeed shown in \cite{nakagawa:molecular} that in the antiparallel configuration the excluded volume is smaller than in the parallel configuration.\footnote{For cylindrically symmetric, convex bodies, possibly tapered along the symmetry axis, it has recently been shown that the antiparallel configuration actually minimizes the excluded volume \cite{palffy:minimum}, a result that had been proved earlier for cones \cite{piastra:octupolar}. How general this property could be for cylindrically, non-convex shapes is still a matter of debate.}

The strategy pursued here is to identify the scalar order parameters of the order tensors in \eqref{eq:probability_density_moments} with the maxima (and conjugated minima) of the density multipoles. From now on, to avoid clutter, we shall drop the subscript $\dens$ from the averages $\aved{\dots}$, as $\dens$ will be the only probability density we shall consider in the following.

\section{Dipolar and Quadrupolar Orders}\label{sec:dipole_and_quadrupole}
Although this paper is concerned with octupolar order in two space dimensions, I find it useful to indulge in rephrasing both dipolar and quadrupolar orders in terms of maxima (and conjugated minima) of the appropriate probability density multipole.

\subsection{Dipole}\label{sec:dipole}
\begin{figure}[h]
\centering
\includegraphics[width=0.25\linewidth]{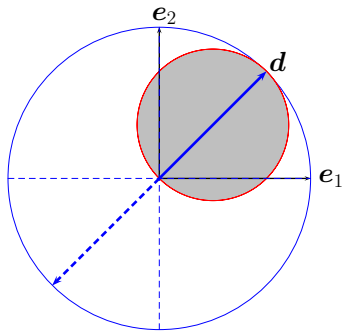}
\caption{(Color online) The dipole $\vd$ superimposed to the polar plot of the normalized dipolar density $\widehat{\rho}_1$ in \eqref{eq:f_1_tilde}, which illustrates pictorially the dipolar probability density. Here $\varphi_0=\frac\pi4$. The vector $-\vd$, which represents the least probable orientation of $\p$, is dashed.}
\label{fig:dials_dipolar}
\end{figure}

The average dipole $\ave{\p}$ is a vector in the plane, which can be represented as
\begin{equation}\label{eq:average_of_p_representation}
\ave{\p}=\lambda_1\vd,
\end{equation}
where $\vd$ is a unit vector and $\lambda_1$, which can be taken as positive, is the \emph{dipolar} scalar order parameter. Letting $\p\cdot\vd=\cos\vartheta$, it follows from \eqref{eq:average_of_p_representation} that $\lambda_1=\ave{\cos\vartheta}$ and so $0\leqq\lambda_1\leqq1$. Alternatively, representing $\p$ in a Cartesian frame $\frame$, as in Fig.~\ref{fig:nail},
\begin{equation}\label{eq:p_representation_phi}
\p=\cos\varphi\,\e_2+\sin\varphi\,\e_1,
\end{equation}
and setting
\begin{equation}\label{eq:d_representation_phi}
\vd:=\cos\varphi_0\,\e_2+\sin\varphi_0\,\e_1,
\end{equation}
we see that
\begin{equation}\label{eq:d_1_d_2_phi}
\lambda_1\sin\varphi_0=\ave{\sin\varphi}\quad\text{and}\quad \lambda_1\cos\varphi_0=\ave{\cos\varphi}.
\end{equation}
Then the function
\begin{equation}\label{eq:f_1}
\rho_1(\p):=\frac1\pi\ave{\p}\cdot\p
\end{equation}
expresses the \emph{dipolar density} in \eqref{eq:probability_density_moments}. Once normalized to its maximum, it simply becomes
\begin{equation}\label{eq:f_1_tilde}
\widehat{\rho}_1(\p)=\vd\cdot\p=\cos(\varphi-\varphi_0)
\end{equation}
and its polar plot,\footnote{The polar plot a scalar function $\rho(\p)$ is the curve described on the plane by the vector $\rho(\p)\p$ as $\p$ ranges over $\disk$.} shown in Fig.~\ref{fig:dials_dipolar}, is a circle passing through the origin.
It should be noted that for $\widehat{\rho}_1(\p)$, as for any \emph{odd} function on $\disk$, the polar plot is actually drawn twice as $\p$ ranges over $\disk$: in one wrapping, $\widehat{\rho}_1$ is positive, whereas it is negative in the other. Thus, the polar plot of $\widehat{\rho}_1$ is just the same as the polar plot of its \emph{positive} part $\widehat{\rho}_1^+:=\max\{0,\widehat{\rho}_1\}$ (or its \emph{negative} part $\widehat{\rho}_1^-:=\min\{0,\widehat{\rho}_1\}$, for that matter).
Clearly, as shown in Fig.~\ref{fig:dials_dipolar}, the dipole $\vd$ corresponds to the direction of $\p$ with maximum dipolar density, $\lambda_1/\pi$, whereas the direction $-\vd$ is corresponds to the direction of minimum dipolar density, $-\lambda_1/\pi$.

\subsection{Quadrupole}\label{sec:quadrupole}
Similarly, the quadrupolar order tensor $\avir{\ptwo}$ is a second-rank tensor which by the Spectral Theorem can be represented as
\begin{equation}\label{eq:average_of_p_o_p_representation}
\avir{\ptwo}=\lambda_2\irr{\etwo},
\end{equation}
where $\e$ is the unit eigenvector of $\avir{\ptwo}$ associated with the positive eigenvalue $\lambda_2$. Letting now $\p\cdot\e=\cos\vartheta$, it follows from \eqref{eq:average_of_p_o_p_representation} that
\begin{equation}\label{eq:lambda_2_representation}
\ave{\cos^2\vartheta}=\lambda_2,
\end{equation}
and so $0\leqq\lambda_2\leqq1$. Setting\footnote{It is perhaps worth mentioning that the angle $\varphi_0$ designating $\e$ through \eqref{eq:quadrupole_e_representation} and featuring in \eqref{eq:a_11_a_12_representation} need \emph{not} be the same as the angle denoted in the same way but featuring in \eqref{eq:d_representation_phi}, as there is no guarantee that $\vd$ and $\e$ should either coincide or be somehow related. This slight abuse of notation is not likely to confuse the reader, if one heeds that only $\vd$ and $\e$ are physically relevant and not the angles that designate them in the plane.}
\begin{equation}\label{eq:quadrupole_e_representation}
\e=\sin\varphi_0\,\e_1+\cos\varphi_0\,\e_2,
\end{equation}
from \eqref{eq:average_of_p_o_p_representation} we also obtain that
\begin{equation}\label{eq:a_11_a_12_representation}
\lambda_2\sin2\varphi_0=\ave{\sin2\varphi}\quad\text{and} \quad\lambda_2\cos2\varphi_0=\ave{\cos2\varphi}.
\end{equation}

\begin{figure}[h]
\centering
\includegraphics[width=0.25\linewidth]{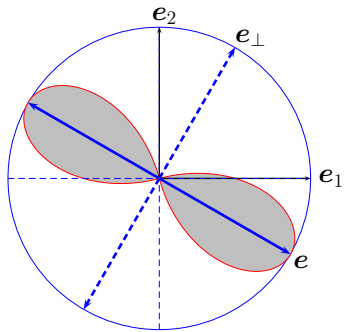}
\caption{(Color online) The polar plot of $\widehat{\rho}_2^+=\max\{0,\widehat{\rho}_2\}$ with $\widehat{\rho}_2$ as in \eqref{eq:f_2_tilde}. Here $\varphi_0=\frac\pi6$. Both unit vectors $\e_\perp$ and $-\e_\perp$ are dashed.}
\label{fig:dials_quadrupolar}
\end{figure}

The mapping defined on $\disk$ by
\begin{equation}\label{eq:f_2}
\rho_2(\p):=\frac2\pi\avir{\ptwo}\cdot\ptwo
\end{equation}
expresses the \emph{quadrupolar density} in \eqref{eq:probability_density_moments}. In complete analogy with $\widehat{\rho}_1$ in \eqref{eq:f_1_tilde}, we write the \emph{normalized} quadrupolar density as
\begin{equation}\label{eq:f_2_tilde}
\widehat{\rho}_2(\p)=2\irr{\etwo}\cdot\ptwo=\cos(2\varphi-2\varphi_0).
\end{equation}
The polar plot of the positive part $\widehat{\rho}_2^+=\max\{0,\widehat{\rho}_2\}$ of $\widehat{\rho}_2$ is depicted in Fig.~\ref{fig:dials_quadrupolar} along with the unit vector $\e$ in \eqref{eq:quadrupole_e_representation} and its orthonormal companion $\e_\perp$. The quadrupolar density $\rho_2$ has equal maxima, $\lambda_2/\pi$, along $\e$ and $-\e$ and equal minima, $-\lambda_2/\pi$, along $\e_\perp$ and $-\e_\perp$.

The rationale behind plotting only the positive part $\widehat{\rho}_2^+$, which is our choice here, is the desire of characterizing the probability density multipoles (at least in two space dimensions) in terms of their maxima (and conjugated minima) and the directions in $\disk$ where they are attained. As will appear clearer in our development below, this is in tune with the notion of generalized eigenvalues and eigenvectors that shall be employed to describe the octupolar order (as well as possibly also higher-rank orders).

\section{Octupolar Order}\label{sec:octupole}
Representing the octupolar order tensor $\avir{\pthree}$ is not as simple as writing the analogue of \eqref{eq:average_of_p_o_p_representation} in the diagonal form
\begin{equation}\label{eq:average_of_p_o_p_o_p_representation_putative}
\avir{\pthree}=\lambda_3\irr{\athree},
\end{equation}
as we lack a generally accepted notion of eigenvalues and eigenvectors for tensors of rank higher than $2$, and, more importantly, for such tensors we lack the analogue of the Spectral Theorem. We shall see below that \eqref{eq:average_of_p_o_p_o_p_representation_putative} can indeed be justified,\footnote{In a way that actually makes the two-dimensional case exceptional.} but this requires some labor and resort to an appropriate notion of generalized eigenvalues and eigenvectors. Here we start from the Cartesian representation
\begin{equation}\label{eq:average_of_p_o_p_o_p_representation_Cartesian}
\avir{\pthree}=a_{ihk}\e_i\otimes\e_h\otimes\e_k,
\end{equation}
where
\begin{equation}\label{eq:a_i_h_k}
a_{ihk}:=\ave{p_ip_hp_k-\frac14\left(p_i\delta_{hk}+p_h\delta_{ik}+p_k\delta_{ih}\right)},
\end{equation}
having denoted by $p_i$ the components of $\p$ in the Cartesian frame $\frame$. It readily follows from \eqref{eq:a_i_h_k} that
\begin{equation}\label{eq:a_symmetries}
\begin{split}
a_{111}+a_{122}&=0,\\
a_{211}+a_{222}&=0,\\
-a_{111}=a_{122}&=a_{221}=a_{212},\\
-a_{222}=a_{211}&=a_{112}=a_{121},
\end{split}
\end{equation}
which show that only two components $a_{ihk}$ are indeed independent; we select $a_1:=a_{111}$ and $a_2:=a_{222}$ to represent all of them. By using again \eqref{eq:p_representation_phi}, we obtain from \eqref{eq:a_i_h_k} that
\begin{equation}\label{eq:a_111_a_222_representation}
a_{111}=-\frac14\ave{\sin3\varphi}\quad\text{and}\quad a_{222}=\frac14\ave{\cos3\varphi}.
\end{equation}

In accord with our earlier treatment of both the dipolar and quadrupolar densities in \eqref{eq:probability_density_moments}, we effectively represent $\avir{\pthree}$ through the maxima (and conjugated, opposite minima) of the function $\rho_3$ defined over $\disk$ by
\begin{equation}\label{eq:f_3}
\rho_3(\p):=\frac4\pi\avir{\pthree}\cdot\pthree,
\end{equation}
which designates  the \emph{octupolar density}.
Looking for the constrained extrema of $\rho_3$ over $\disk$ amounts to solving the problem
\begin{subequations}\label{eq:eigenvalue_problem}
\begin{equation}\label{eq:stationarity_condition}
a_{ihk}x_hx_k=\lambda x_i,\quad i=1,2,
\end{equation}
for $\x\in\disk$ ad $\lambda\in\real$, where $\lambda$ is the Lagrange multiplier associated with the constraint
\begin{equation}\label{eq:x_constraint}
x_1^2+x_2^2=1.
\end{equation}
\end{subequations}
The solutions $(\lambda,\x)$ of problem \eqref{eq:eigenvalue_problem}  coincide with the generalized eigenvalues and eigenvectors of $\avir{\pthree}$ according to a definition introduced  for tensors (not necessarily symmetric) of rank higher than $2$ in finite-dimensional spaces of any dimension. This concept has been proposed and made precise in \cite{qi:eigenvalues,qi:eigenvalues_2007,ni:degree}. Different notions of generalized eigenvalues and eigenvectors have been introduced in the literature; for the one chosen here, a theorem was recently proved in \cite{cartwright:number} on the cardinality of the eigenvalues, which to my knowledge is not available for other notions. According to this theorem, in the case at hand there are at most \emph{three} distinct complex eigenvalues (defined to within a sign), of which at least one is real. We shall show below that the solutions of \eqref{eq:eigenvalue_problem} are rather easy to find: they turn out to be all real and equal to one another, to within a sign.

\begin{figure}[h]
\centering
\includegraphics[width=0.5\linewidth]{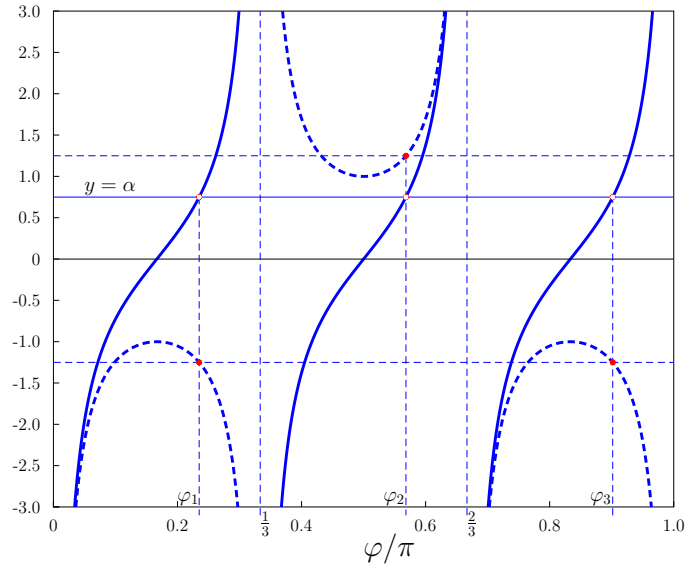}
\caption{(Color online) Graphical illustration of the solutions of \eqref{eq:phi_equations} for $a_1\neq0$. The $\varphi_i$'s, which are delivered by the intersections (hollow circles) of $y=\alpha$ (solid thin line) with $y=-1/\tan3\varphi$ (solid thick lines), differ by $\frac\pi3$ from one another. The corresponding $\lambda^{(i)}$, which are read off from the graph of $\Lambda(\varphi)=-1/\sin3\varphi$ (full circles on the thick dashed lines), alternate in sign. Here $\alpha=\frac34$ and $|\lambda^{(i)}|=\frac54|a_1|$.}
\label{fig:plots}
\end{figure}

By use of \eqref{eq:a_symmetries}, we write \eqref{eq:stationarity_condition} in the explicit form
\begin{subequations}\label{eq:x_equations}
\begin{align}
a_1(x_1^2-x_2^2)-2a_2x_1x_2&=\lambda x_1,\label{eq:x_1_equation}\\
a_2(x_2^2-x_1^2)-2a_1x_1x_2&=\lambda x_2,\label{eq:x_2_equation}
\end{align}
\end{subequations}
which make it evident that to each solution $(\lambda,\x)$ there corresponds the solution $(-\lambda,-\x)$. To solve \eqref{eq:x_equations} subject to \eqref{eq:x_constraint} we parameterize the latter by letting, as in \eqref{eq:p_representation_phi},
\begin{equation}\label{eq:x_parameterization}
x_2=\cos\varphi\quad\text{and}\quad x_1=\sin\varphi.
\end{equation}
The parity symmetry $(\lambda,\x)\mapsto(-\lambda,-\x)$ thus translates into $(\lambda,\varphi)\mapsto(-\lambda,\varphi+\pi)$, so that each solution $\varphi\in[0,\pi]$ generates another solution in $[\pi,2\pi]$ by a $\pi$-shift. Inserting \eqref{eq:x_parameterization} into \eqref{eq:x_equations}, we arrive at
\begin{subequations}\label{eq:phi_equations}
\begin{align}
-a_1\cos2\varphi-a_2\sin2\varphi&=\lambda\sin\varphi,\label{eq:phi_1_equation}\\
a_2\cos2\varphi-a_1\sin2\varphi&=\lambda\cos\varphi.\label{eq:phi_1_equation}
\end{align}
\end{subequations}
We distinguish two cases, $a_1=0$ and $a_1\neq0$. In the former case, eliminating $\lambda$ from \eqref{eq:phi_equations} we obtain
\begin{equation}\label{eq:a_1_0}
\tan2\varphi+\tan\varphi=0,
\end{equation}
which in $[0,\pi]$ has solutions $\varphi_1=0$, $\varphi_2=\frac\pi3$, and $\varphi_3=\frac23\pi$. Correspondingly, $\lambda$ is delivered by $\lambda^{(1)}=a_2$, $\lambda^{(2)}=-a_2$, and $\lambda^{(3)}=a_2$. If $a_1\neq0$, we set $\alpha:=\frac{a_2}{a_1}$ and eliminating again $\lambda$ we obtain
\begin{equation}\label{eq:alpha_equation}
\alpha=-\frac{1}{\tan3\varphi},
\end{equation}
which has three solutions in $[0,\pi]$, denoted $\varphi_1$, $\varphi_2$, and $\varphi_3$, to which there correspond three values of $\lambda$, $\lambda^{(i)}=a_1\Lambda(\varphi_i)$, delivered by the function
\begin{equation}\label{eq:Lambda}
\Lambda(\varphi):=-\frac{1}{\sin3\varphi}.
\end{equation}
The strategy to solve \eqref{eq:phi_equations} for $a_1\neq0$ is illustrated graphically in Fig.~\ref{fig:plots}.
For a generic $\alpha$, the $\varphi_i$'s are obtained by intersecting the line $y=\alpha$ with the graph of $y=-1/\tan3\varphi$. Correspondingly, the $\lambda^{(i)}$ (scaled to $a_1$) are read off from the graph of $y=\Lambda(\varphi)$. Since $\Lambda(\varphi\pm\frac\pi3)=-\Lambda(\varphi)$, and any two adjacent roots $\varphi_i$ differ by $\frac\pi3$, the corresponding $\lambda^{(i)}$ alternate in sign and for all of them $|\lambda^{(i)}|=|a_1|\sqrt{1+\alpha^2}$. The solution shown in Fig.~\ref{fig:plots} is for $\alpha>0$ and delivers $\lambda^{(1)}<0$ if also $a_1>0$. For $\alpha<0$ and $a_1>0$, $\lambda^{(1)}$ would also change its sign.

\begin{figure}[h]
\centering
\includegraphics[width=0.25\linewidth]{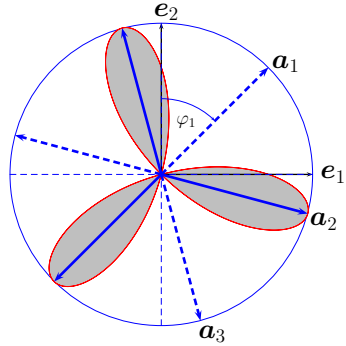}
\caption{(Color online) The $6$ unit vectors $\va_i$, conjugated by parity, that correspond to the solutions of \eqref{eq:x_equations} shown in Fig.~\ref{fig:plots}. The dashed vectors $\va_1$, $-\va_2$, and $\va_3$, which correspond to the eigenvectors of the tensor in \eqref{eq:average_of_p_o_p_o_p_representation_Cartesian} with negative generalized eigenvalues, represent the orientations of $\p$ with the least octupolar probability density $\rho_3$. The superimposed polar plot represents the positive part $\widehat{\rho}_3^+$ of the normalized octupolar density $\widehat{\rho}_3$.}
\label{fig:dials}
\end{figure}

It is worth noting that all extrema of $\rho_3$ on $\disk$ are either maxima or minima (conjugated by parity to the former): no saddle thus corresponds to a generalized eigenpair of $\avir{\pthree}$. The eigenvectors $\va_i=x^{(i)}_h\e_h$ of $\avir{\pthree}$ are then $6$, counting all solutions of \eqref{eq:x_equations} conjugated by parity. They will be conventionally represented by \emph{solid} vectors if corresponding to \emph{positive} eigenvalues and by \emph{dashed} vectors if corresponding to \emph{negative} eigenvalues. The solutions shown in Fig.~\ref{fig:plots} are reproduced in Fig.~\ref{fig:dials} with this convention.

Rescaling the octupolar density $\rho_3$ in \eqref{eq:f_3} to its maximum, we reduce it to a simple function of $\varphi$, $\widehat{\rho}_3=\cos(3\varphi-3\varphi_0)$, where $\varphi_0$ is any solution of \eqref{eq:alpha_equation}.\footnote{Again, I am guilty here of a slight abuse of notation, as $\varphi_0$ has already been used above to represent $\vd$ and $\e$, and the corresponding normalized densities $\widehat{\rho}_1$ and $\widehat{\rho}_2$. No relation should be expected between these angles, though they are occasionally denoted in the same way.} Compared to the polar plots of $\widehat{\rho}_1$ and $\widehat{\rho}_2^+$ in \eqref{eq:f_1_tilde} and \eqref{eq:f_2_tilde}, which exhibit one and two lobes respectively, the polar plot of $\widehat{\rho}_3^+:=\max\{0,\widehat{\rho}_3\}$, the positive part of $\widehat{\rho}_3$, displays one more, as expected; in Fig.~\ref{fig:dials}, it is superimposed on the $6$ dials that represent the eigenvectors $\va_i$.

Here we have chosen to characterize the third-rank tensor $\avir{\pthree}$ in terms of its generalized eigenvalues and eigenvectors corresponding to maxima and minima of $\rho_3$. However, in the two-dimensional setting, equation \eqref{eq:average_of_p_o_p_o_p_representation_putative} is also proved valid letting $\va$ be \emph{any} generalized eigenvector of $\avir{\pthree}$ and $\lambda_3=4\lambda$, where $\lambda$ is the eigenvalue associated with the selected $\va$. To see this, it suffices to represent $\va$ in \eqref{eq:average_of_p_o_p_o_p_representation_putative} as
\begin{equation}\label{eq:a_phi_representation}
\va=\cos\varphi\,\e_2+\sin\varphi\,\e_1,
\end{equation}
which leads us to identify the two independent components of $\avir{\pthree}$ as
\begin{subequations}\label{eq:a_1_a_2_extra_representation}
\begin{align}
a_1&=\lambda_3\sin\varphi\left(\sin^2\varphi-\frax34\right),\label{eq:a_1_extra_representation}\\
a_2&=\lambda_3\cos\varphi\left(\cos^2\varphi-\frax34\right).\label{eq:a_1_extra_representation}
\end{align}
\end{subequations}
Solving these equations for $\lambda_3$ and $\varphi$, we readily obtain that $\lambda_3=4a_1\Lambda(\varphi)$, where $\Lambda(\varphi)$ is as in \eqref{eq:Lambda}, and $\varphi$ is a root of \eqref{eq:alpha_equation}.
We thus conclude that in the two-dimensional case the generalized eigenpairs $(\lambda,\va)$ of $\avir{\pthree}$ afford a $6$-fold degenerate representation of this third-rank tensor through \eqref{eq:average_of_p_o_p_o_p_representation_putative}, in complete analogy to what the Spectral Theorem does for a second-rank tensor.\footnote{In particular, this result fully justifies the representation in equations (44) of \cite{ohta:deformation} and (A4) of \cite{tarama:oscillatory} for the third-rank shape tensor adopted there to describe a self-propelled cell in two space dimensions.} It will be shown in \cite{gaeta:octupolar} that this is indeed a rather exceptional circumstance.

\section{Conclusions}\label{sec:conclusion}
The strategy inspiring our quest for the representation of the octupolar order in any space dimension can be easily summarized by saying that no diagonalization need in general be attempted for a third-rank fully symmetric and traceless tensor (an octupolar tensor, for short). This latter should rather be characterized in terms of the maxima (and conjugated, opposite minima) of the octupolar probability density. In algebraic terms, this amounts to compute the relevant generalized eigenvalues and eigenvectors according to a definition for which a theorem concerning their cardinality has recently been proved \cite{cartwright:number}.

In two space dimensions, an octupolar tensor is simply described by $2$ scalar parameters, and so its generalized eigenvalues and eigenvectors in the plane must be so constrained as to be described by $2$ parameters only. It was indeed shown here that in the two-dimensional case all generalized eigenvalues of an octupolar tensor are equal (to within a sign) and that its three inequivalent eigenvectors are described by a single rotation angle. It will be shown in \cite{gaeta:octupolar} how the $7$ independent parameters that describe an octupolar tensor in three space dimensions concur to either the \emph{three} or  \emph{four} inequivalent generalized eigenvalues and eigenvectors which are associated in the generic case with the maxima of the octupolar probability density and the directions along which they are attained.

A critical voice could be raised against the mathematical machinery devised here to describe the multipoles of the probability density $\dens$ in \eqref{eq:probability_density_moments}. Comparing $\widehat{\rho}_1$, $\widehat{\rho}_2$, and $\widehat{\rho}_3$ above,\footnote{See, for example, \eqref{eq:f_1_tilde} and \eqref{eq:f_2_tilde}.} one could easily argue that the multipoles in \eqref{eq:probability_density_moments} are nothing but the Fourier components of $\dens$ in $\disk$. Such a critique could indeed be grounded if the angle $\varphi_0$ were indeed the same in all the formulae where it appears, which would the case if the vectors $\vd$, $\e$, and $\va$ were somehow \emph{locked}, that is, rigidly related to one another, which they are \emph{not}. In general, there is no reason (apart from convenience and laziness) why one should think that dipolar, quadrupolar, and octupolar order tensors have correlated  eigenvectors.

In two space dimensions, a \emph{diagonalized} form for an octupolar tensor was established in \eqref{eq:average_of_p_o_p_o_p_representation_putative}, which unfortunately has no analogue in three space dimensions \cite{gaeta:octupolar}. It is thus conceivable that the method proposed here to identify the scalar order parameters of an octupolar tensor could be extended to any space dimension, whereas the diagonal form in \eqref{eq:average_of_p_o_p_o_p_representation_putative} is only accidentally valid in two space dimensions.

\begin{acknowledgements}
I was introduced to the problem of representing higher-rank order tensors by Peter Palffy-Muhoray; discussions we had on this topic (mainly at lunchtime) while we were both Visiting Fellows at the \emph{Isaac Newton Institute} in Cambridge during the Programme on \emph{The Mathematics of Liquid Crystals} are gratefully acknowledged. I am also indebted to Stephen Morris for showing his captivating experiments to me. I hope that the representation of octupolar order proposed here could indeed be useful to describe quantitatively his data.
\end{acknowledgements}


%

\end{document}